%% file: main.tex
\documentclass{article}
\usepackage{graphicx} % Required for inserting images
\usepackage{listings}
\usepackage{courier}
\usepackage{siunitx}
\usepackage{listings}
\usepackage{hyperref}
\usepackage{nomencl} % If needed
\makenomenclature
\usepackage{amsmath}
\usepackage{algorithm2e}
\usepackage{bm}
\usepackage{booktabs}

\usepackage[letterpaper, total={6in, 9in}]{geometry}

\title{Performance Portable Monte Carlo Neutron Transport in MCDC via Numba\footnote{
This is an Accepted Manuscript of an article published by IEEE in Computing and Science and Engineering (CISE) on March 14 2025, available at: \url{https://doi.org/10.1109/MCSE.2025.3550863}} \footnote{
Please cite as: J. P. Morgan, I. Variansyah, B. Cuneo, T. S. Palmer and K. E. Niemeyer, "Performant and Portable Monte Carlo Neutron Transport via Numba," in Computing in Science \& Engineering, vol. 27, no. 1, pp. 57-65, Jan.-March 2025, doi: \href{https://doi.org/10.1109/MCSE.2025.3550863}{10.1109/MCSE.2025.3550863}. }
}
\date{%
    \small{
    $^1$Center for Exascale Monte Carlo Neutron Transport (CEMeNT)\footnote{https://cement-psaap.github.io/}\\
    $^2$School of Mechanical Industrial and Manufacturing Engineering, Oregon State University\\%
    $^3$School of Nuclear Science and Engineering, Oregon State University\\%
    $^4$Department of Computer Science, Seattle University%
    }
}

\author{
  Joanna Piper Morgan$^{1,2}$\footnote{morgajoa@oregonstate.edu, joannapipermorgan@gmail.com}
  \and
  Ilham Variansyah$^{1,3}$ \footnote{variansi@oregonstate.edu}
  \and
  Braxton Cuneo$^{1,4}$
  \and
  Todd S. Palmer$^{1,3}$
  \and
  Kyle E. Niemeyer$^{1,2}$
}

\begin{document}

\maketitle

\begin{abstract}\looseness-1
Finding a software engineering approach that allows for portability, rapid development, and open collaboration for high-performance computing on GPUs and CPUs is a challenge. 
We implement a portability scheme using the Numba compiler for Python in Monte Carlo / Dynamic Code (MC/DC), a new neutron transport application for rapidly developing Monte Carlo. 
Using this scheme, we have built MC/DC as an application that can run as a pure Python, compiled CPU, or compiled GPU solver. 
In GPU mode, we use Numba paired with an asynchronous GPU scheduler called Harmonize to increase GPU performance. We present performance results (including weak scaling up to 256 nodes) for a time-dependent problem on both CPUs and GPUs and compare favorably to a production C++ code.
\end{abstract}

\hfill

\input{full_text}

\bibliographystyle{IEEEtran}
\bibliography{references}

\end{document}

%% file: full_text.tex
% the actual text of the article is here for ease of porting from format to format

%if we are doing this in CiSE we will only have 12 citations!

Developing software to simulate physical problems that demand high- performance computing (HPC) is difficult.
Modern HPC systems commonly use both CPUs and GPUs from various vendors.
Years can be spent porting a code from CPUs to run on GPUs, then again when moving from one GPU vendor to the next \cite{pozulp_progress_2023}.

Portability issues compound when designing software for rapidly developing numerical methods where algorithms need to be both implemented and tested at scale.
Finding a software engineering approach that balances the need for portability, rapid development, open collaboration, and performance can be challenging especially when numerical schemes do not rely on operations typically implemented in libraries (i.e., linear algebra as in LAPACK or Intel MKL). 
% one paragraph for CISE reqirments

HPC software engineering requirements can be met using a Python-as-glue-based approach, where peripheral functionality (e.g., MPI calls, I/O) is implemented using Python packages but compiled functions are called through Python's C-interface where performance is needed.
Python-as-glue does not necessarily assist in producing the compiled compute kernels themselves---what the Python is gluing together---but can go a long way in simplifying the overhead of peripheral requirements of HPC software.
With this technique, environment management and packaging uses \texttt{pip}, \texttt{conda}, or \texttt{spack}, input/output with \texttt{h5py}, MPI calls with \texttt{mpi4py}, 
and automated testing with \texttt{pytest}, which can all ease initial development and continued support for these imperative operations. 

Many tools have been developed to extend the Python-as-glue scheme to allow producing mostly single-source compute kernels for both CPUs and GPUs.
% a DSL, pyfr
One tactic is using a domain-specific language to avoid needing a low-level language (e.g., FORTRAN, C).
A domain-specific language is designed to alleviate development difficulties for a group of subject-area experts and can abstract hardware targets if defined with that goal.
%It can even abstract hardware targets if it is defined with that goal.
PyFR, for example, is an open-source computational fluid dynamics solver that implements a domain-specific language plus Python structure to run on CPUs and Nvidia, Intel, and AMD GPUs~\cite{pyfrPetascale}. 
%The overhead of this Python glue is less than 1\% in PyFR.
Witherden et al.~\cite{pyfrPetascale} discussed how this scheme allows PyFR developers to rapidly deploy numerical methods at deployment HPC scales and have demonstrated performance at the petascale.

Other projects have addressed the need to write user-defined compute kernels entirely in Python script.
Numba is a compiler that lowers a small subset of Python code with NumPy arrays and functions into LLVM, then just in time (JIT) compiles to a specific hardware target \cite{lam_numba_2015}. 
Numba can also compile global and device functions for Nvidia GPUs from compute kernels defined in Python.
API calls are made through Numba on both the Python side (e.g., allocate and move data to and from the GPU) and within compiled device functions (e.g., to execute atomic operations).

When compiling to GPUs, Numba supports an even smaller subset of Python, losing most of the operability with NumPy functions.
If functions are defined using only that smallest subset, Numba can compile the same functions to CPUs or GPUs, or execute those functions purely in Python.
Numba data allocations on the GPU can be consumed and ingested by functions from CuPy if linear-algebra operations are required in conjunction with user-defined compute kernels.

Motivated by Numba's capabilities, we developed a new Monte Carlo neutron transport application called Monte Carlo / Dynamic Code\footnote{\url{https://github.com/CEMeNT-PSAAP/MCDC}} (MC/DC) \cite{mcdc_joss_2024, variansyah_mc23_mcdc}.
Our development of MC/DC uses a Numba+Python development scheme along with a GPU event scheduler to abate portability issues and allow for rapidly developing novel numerical methods at the HPC scale on CPUs and GPUs.

In this paper, we first introduce neutron transport and the Monte Carlo solution method.
We next describe in greater detail MC/DC's novel (for the field) software engineering approach on CPUs and GPUs, along with pitfalls and difficulties when using this development scheme.
We discuss how novel numerical methods are prototyped and developed in MC/DC, and supported for execution on both GPUs and CPUs.
Then, we analyze the compute performance of MC/DC and, where possible, compare it against modern production Monte Carlo neutron transport solvers.
Finally, we provide concluding remarks and outline future work.

% how'd we get to python

%divergent bakends code base per hardware target written in multiple languages.

%Numba has been shown to be slower then other high level portability frameworks for unoptimized matrix multiplication \cite{Godoy_2023}.
%Monte Carlo neutronic workflows are so memory bound that it's doubtful even significant changes to FLOP performance of a 

\section{Monte Carlo Neutron Transport and MC/DC}

Predicting how neutrons move through space and time is important when modeling inertial confinement fusion systems, pulsed neutron sources, and nuclear criticality safety experiments, among other systems.
Unlike charged particles, neutrons can interact with the nucleus of an atom because they are unaffected by the negatively charged orbital electrons and the positively charged core.
Some isotopes readily absorb neutrons into the nucleus, which may make such atoms unstable.
When an unstable atom fissions, it releases energy along with two daughter nuclei and subatomic particles, which may be more neutrons, depending on the parent atom.
If additional neutrons are released and encounter more fissionable material, the release of subsequent neutrons can induce a chain reaction.
Thus, a population of neutrons can change rapidly in dynamic systems.

Simulating neutron transport problems is computationally difficult using any numerical method, because the neutron distribution is a function of seven independent variables: three in space, three in velocity, and time \cite{lewis_computational_1984}.
% This problem demands HPCs and specifically the exascale
Modern HPC systems now enable high-fidelity simulation of neutron transport for problem types that have seldom been modeled before due to limitations of previous computers. % need citation?
Specifically, large-scale, highly dynamic transport problems require thousands of compute nodes using modern hardware accelerators (i.e., GPUs) \cite{hamilton_continuous-energy_2019, romano_openmc_nodate}.

The behavior of neutrons can be modeled with a Monte Carlo simulation, where particles with statistical importance are created and transported to produce a particle history \cite{lewis_computational_1984}.
A particle's path and the specific set of events that occur within its history are governed by pseudorandom numbers, known probabilities (e.g., from material data), and known geometries.
Data about how particles move and/or interact with the system are tallied to solve for parameters of interest with an associated statistical error from the Monte Carlo process.

The analog Monte Carlo method converges slowly at a rate of $\mathcal{O}(1/\sqrt{n})$, where $n$ is the number of simulated particles.
New Monte Carlo schemes could converge the solution faster in wall-clock time with fewer simulated particles and may be needed to effectively simulate some systems.
We wrote MC/DC to enable rapidly developing these novel numerical methods for time-dependent simulations in particular.

%We conducted an initial set of investigations to compare what we considered to be the most viable options at the time. 
%We compared the same algorithm in various implementations using PyKokkos, PyCUDA/PyOpenCL, and Numba \cite{morgan2022}.
%We found that all three methods produced similar runtimes for our workflows on CPUs and GPUs for a simple transient Monte Carlo neutron transport simulation.
%Ultimately, we decided to use a Numba + mpi4py development scheme to build out a Monte Carlo neutron transport code for rapid numerical methods development, portable to various HPC architectures.

\begin{figure*}
    \centerline{\includegraphics[width=.95\textwidth]{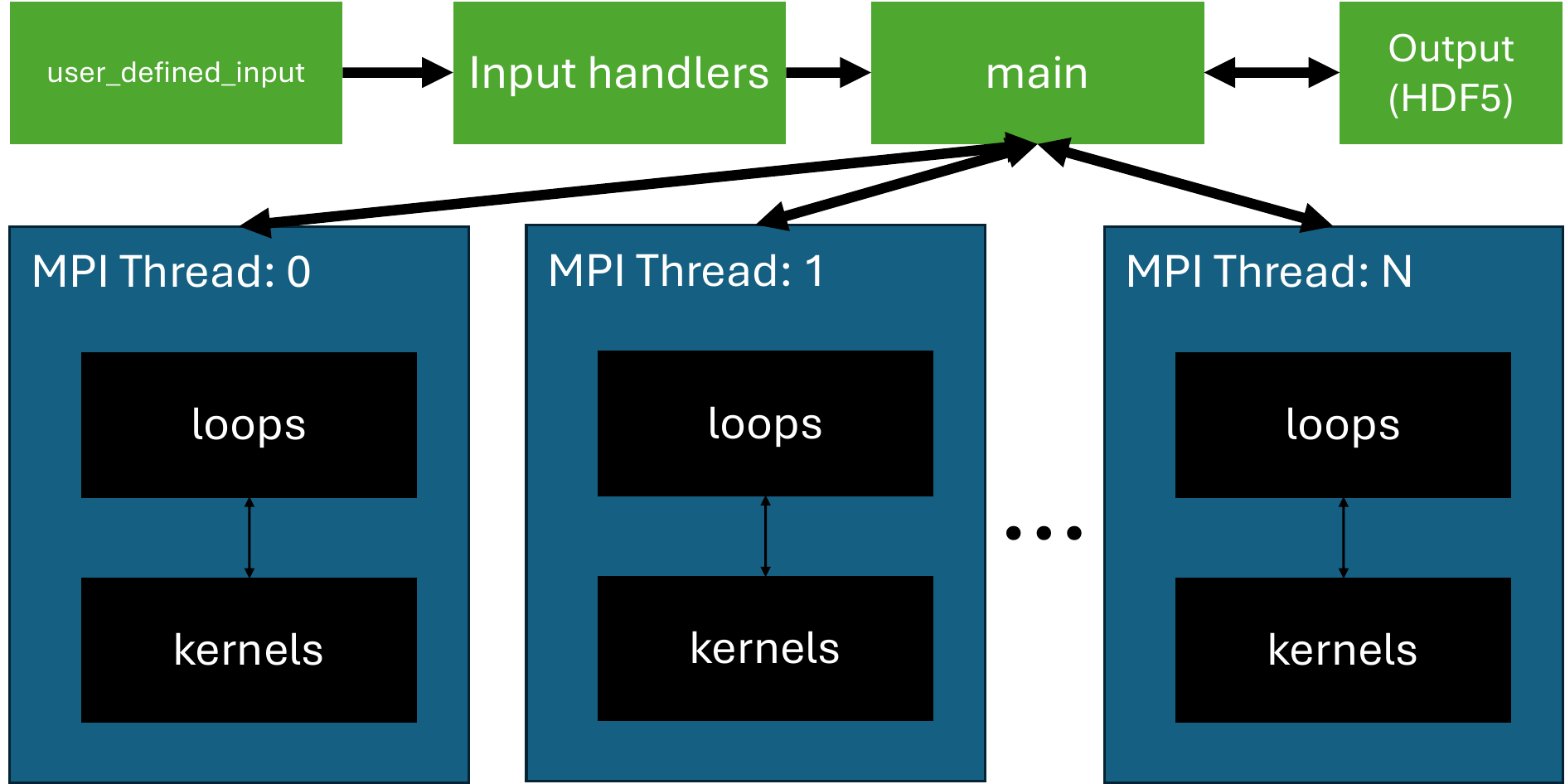}}
    \caption{MC/DC's overall structure and how functions get called and interact. Green functions are entirely Python, and black functions are compiled compute kernels if they are running in Numba CPU or GPU modes.}\vspace*{-5pt}
    \label{mpi_mcdc}
\end{figure*}

MC/DC offers a similar feature set as other Monte Carlo neutron transport applications (e.g., OpenMC \cite{romano_openmc_nodate}, Shift \cite{hamilton_continuous-energy_2019}) with support for $k$-eigenvalue and fully time-dependent simulation modes in full three-dimensional constructive solid geometry.
It can model the neutron distribution in energy using either continuous energy or multigroup nuclear data.
It also supports domain decomposition.
All these features are supported on CPU (x86, ARM, POWERPC) and GPU processor targets (Nvidia and AMD), with MPI to target multiple processors (via mpi4py \cite{dalcin_mpi4py_2021}).

The number of novel schemes and simulation techniques implemented in MC/DC in a short time illustrates the success in its software engineering structure.
MC/DC supports the use of the iterative quasi-Monte Carlo (iQMC) method where deterministic and Monte Carlo transport operations run in tandem to converge solutions faster than they would in a pure Monte Carlo method. %\cite{pasmann_iqmc_nodate}
Other novel developments include global sensitivity analysis, hash-based random number generation for fully replicable solution testing, time-dependent population control, and continuously moving surfaces.
Several ongoing developments include quasi Monte Carlo, residual Monte Carlo, and machine learning techniques for dynamic node scheduling.

\section{MC/DC on CPUs}

To compile on CPUs, MC/DC uses the Numba compiler for Python to lower compute functions into LLVM and compile for a specific hardware target.
Figure~\ref{mpi_mcdc} shows MC/DC's functional layout when running in both MPI and Numba mode.

First, a user writes a Python script and imports \texttt{mcdc} as a package and forms an input script.
Then, the user interfaces with functions described in the input handler within MC/DC describing the physical models, material data, and simulation parameters.
This input layout is similar to how other Monte Carlo neutron transport applications input problems.

The input script calls a run command, which starts initialization functions within MC/DC. 
The initialization process allocates and constructs a global variable containing the user-defined inputs, meshes, particle banks, event tallies, and current global states.
This global variable is formed from a statically typed NumPy \texttt{ndarray}, which acts like a Python dictionary, where keywords are used to extract numerical arrays.
After building the global variable, initialization functions dispatch MPI processes if running in MPI mode, and begin the Monte Carlo neutron transport simulation.

Each MPI process calls functions containing the various transport algorithms and modes that MC/DC supports.
Each transport function is decorated with a Numba JIT (\texttt{@jit}) compilation flag declaring that each function must be compiled before being executed if running in Numba mode.
These transport logic loops are the highest level at which Python functions will be compiled in MC/DC.
For example, a fixed source problem will loop over all the particles and transport them until the particle's history is terminated from a physical event (e.g., capture, fission, time or space boundary), a simulation event (e.g., time census), or a variance-reduction event (e.g., population control, implicit capture).

The specific functions within each algorithm that conduct the actual transport operations (e.g., moving particles, tallying events, generating daughter particles from fission) are contained in the kernel set where all functions are \texttt{@jit} decorated.
Figure~\ref{fig:jitfunctions} shows an example compute kernel that updates the position and time of a particle as it moves.
It also shows the declaration of an \texttt{numpy.ndarray} data structure used in MC/DC.

\begin{figure}
\begin{lstlisting}[language=Python]
import numpy
from numba import jit

part = numpy.dtype([
    ('x', float64), ('y', float64),
    ('z', float64), ('ux', float64),
    ('uy', float64), ('uz', float64),
    ('v', float64) ])

@jit
def move_particle(P: part, distance):
    P['x'] += P['ux'] * distance
    P['y'] += P['uy'] * distance
    P['z'] += P['uz'] * distance
    P['t'] += distance / P['v']
\end{lstlisting}
\caption{Example of a decorated function and MC/DC's data structures based on \texttt{numpy.ndarray}.}
\label{fig:jitfunctions}
\end{figure}

After all transport is completed and the simulation is finished, the program returns to the Python interpreter and calls finalization functions.
Here, requested tally information along with statistical error provided from the Monte Carlo process are saved in an HDF5 file.
Data can be extracted from this HDF5 file and used in Python scripts to do post-processing analysis and/or data visualization with tools like Matplotlib, or post-processing can be done in other applications like Visit or Paraview.

%On a single CPU the total number of threads that can be ruing simultaneously generally corresponds to the number of cores\footnote{
%    Many modern CPUs have more ``logical" cores than physical cores through the use simultaneous multi-threading or ''hyper-threading". They also may or may not have additional vector lanes per-core to take advantage of single instruction-multiple data type-parallelism (e.g., using SSE or AVX instructions).
%}.
%When running the same application different threads can take completely different paths in a program with little impact to performance.
%These can be targeted with either shared memory parallelism techniques (e.g., using tools like Pthreads or OpenMP) or distributed memory parallelism (e.g., MPI).
%The latter of which allows for the targeting of not only multiple cores on a single CPU die, but also multiple CPUs on completely different nodes of an HPC.

When initially exploring a novel transport method, a developer can work in a pure Python environment where functions are entirely executed in the Python interpreter.
In this mode, the developer can bring any package into any function, do typing dynamically, and use any Python data structure.
MC/DC can be executed in MPI mode in Python as well as compiled CPU mode.
While a full Python development environment is great for initially proving a concept, it often proves to be too slow for problems of interest.

When more performance is required, developers rewrite their kernels to strictly use Numba-enabled functions.
Numba only supports a small subset of the Python ecosystem. 
Some Python data structures like dictionaries and lists can no longer be used and must instead come from NumPy implementations. 
Thus, when using Numba, the small subset of functions supported effectively becomes a domain-specific language.

Scientific computing using Python is often done with NumPy functions and data structures, making these fairly natural for numerical-methods developers to use and understand.
In fact, we have found NumPy functionality to be more commonly used in initial development than other non-supported Python methods, making the restrictions in Numba more palatable.
Some developers report skipping Python-mode development entirely and starting with Numba-CPU work for their initial proofs of concept, as they find that aids in future debugging efforts.
Similarly, other developers report making small, incremental changes in Python-based algorithms, then checking to ensure successful compilation in Numba before moving forward, roughly at every commit.
When kernels are written to support Numba mode, they can be compiled to any supported CPU targets automatically (i.e., x86, ARM64, PPC64).

We can identify pitfalls with this approach, the most significant of which are:
\begin{itemize}
    \item Common failures of \texttt{numba.object\_mode},
    \item Lack of MPI calls from within the JIT-compiled Numba code,
    \item Numba kernel debugging and profiling;
    \item Loss of functions from SciPy not implemented in NumPy, and
    \item Restrictions with \texttt{numpy.ndarray} as our primary data structure.
\end{itemize}
Most of these issues have workarounds, but make implementing numerical methods in Numba harder.

Consider that \texttt{numpy.ndarray} requires ``square'' size allocation for all elements such that the size of every named element within an array must be the same.
If one element requires \num{10000} data points and the next only \num{100}, the size of that \texttt{numpy.ndarray} is \num{20000}, which is a drastic over-allocation.
This is a primary issue for continuous energy material data, where some materials may require tens of thousands of points to fully resolve, and others may only need hundreds.
While Numba does have some features to help in this circumstance (namely experimental \texttt{jit\_classes}), we must keep the \texttt{numpy.ndarray} to support MPI calls and GPU portability.
To fix this issue, given our constraints, we are moving towards using one-dimensional vectors with length information to offset between different variables, potentially impacting MC/DC's developer-friendliness.
%This very primitive memory technique is similar to how data might be stored in low-level languages like FORTRAN.
Accepting increased complexity to achieve portability is common in MC/DC, so developing in it can be about as difficult as in a low-level language.
%For another example: mpi4py calls are not supported from within Numba compiled kernels.
%Numba offers a function called \texttt{object\_mode} to allow for non-copmilable Python features to be accessed from Python code.
%However we have found \texttt{object\_mode} to create significant debugging issues with very little compiler messaging as to what has gone wrong.
%Furthermore, \texttt{object\_mode} can significantly impact performance of executing kernels making it a last-resort for bringing non-Numba methods into Numba kernels.

Other deficiencies are known to the Numba community, and some even have ongoing open-source remedies.
For example, \texttt{numba-mpi}\footnote{\url{https://github.com/numba-mpi/numba-mpi/}} is a project to support compiled-side MPI calls, \texttt{Profila}\footnote{\url{https://github.com/pythonspeed/profila}} attempts to bring the GNU debugger to Numba kernels, and \texttt{numba-scipy}\footnote{\url{https://github.com/numba/numba-scipy}} extends support for more SciPy functions to Numba.
However, most of these community projects are still in their infancy and not robust enough to handle the large and complex structures in MC/DC.
%\revise{Developing in Numba has improved in the years since work in MC/DC began in 2021. Specifically error reporting and debug messages have seen massive improvements in Numba version $\geq$ 0.59.0.
%While these error messages are an improvement they are not a replacement for a debugger.}{editor}

For CPU-based HPC deployments, a Python-as-glue strategy with Numba compute kernels can enable portable (between CPU architectures and scales) and high-performance code.
%While it is about as difficult to develop using Numba+Python code for complex algorithms the same Numba kernels that where developed for CPU based implementations can hypothetically be compiled to the GPU.
However, on GPUs, if using Python+Numba alone, a developer must still have in-depth understanding of their target GPU parallelism paradigm to achieve high performance.

\section{MC/DC on GPUs}

GPUs use a single-instruction multiple-thread (SIMT) parallelism paradigm, where threads are executed in teams called warps, or wavefronts, and do the same operations in lockstep. 
If threads in the same warp need to take different paths in a program (e.g., different if/else branches or iterating loops a different number of times), each path must be executed serially.
This behavior is called thread divergence.
Threads that do not belong to the currently executing path are disabled so that the end result of the computation is consistent with the control flow logic.
Mitigating thread divergence will usually result in higher performance of GPU-enabled applications.

Unfortunately, commonly implemented Monte Carlo neutron-transport algorithms are examples of highly divergent workflows, as the behavior of any individual particle is governed by random numbers.
Much more work is often required beyond naive syntax porting to implement Monte Carlo radiation transport applications to GPUs \cite{pozulp_progress_2023}.
%A single GPU can have be hundreds, even thousands of warps, each with tens or hundreds of threads which is why (for algorithms that can take advantage of there parallelism scheme) they are so performant.

%\revise{
%While a Python+Numba scheme can alleviate difficulties with GPU kernel production, it is not sufficient on it's own to enable algorithms for CPU 
%to run performantly under SIMT execution paradigm, or assisting with more complex GPU memory management (both between CPU and GPU and on the GPU between individual warps).
%MC/DC's ultimate goal is to divorce the numerical methods development from the software engineering structure so when moving to deploy on GPUs devising an approach that can treat and abstract these two issues is needed.}{revTwo}
%While MC/DC does implement some specific algorithms designed to increase GPU performance (e.g. event-based algorithms) more is needed to enable a Monte Carlo neutron transport application to be performant on GPUs.
%Someone, somewhere is still gonna have to do the dirty-work and put their hands in filth@braxton.

%\revise{The goal of MC/DC is to abstract the target architecture from numerical methods developer as much as possible.
%Numba alone does not sufficiently do this for Monte Carlo logic as algorithms must still be written to explicitly enable the SIMT programming model GPUs employ.}{revTwo}
% cite the harmonize zenodo repo before publication
When compiling and running on GPUs, MC/DC uses an open-source asynchronous event scheduling library called Harmonize\footnote{\url{https://github.com/CEMeNT-PSAAP/harmonize}}~\cite{brax2023} to 
reorganize the execution of business logic and storage/movement of data to better fit the SIMT execution paradigm of GPUs.
Harmonize implements runtimes that examine operations due to be executed, segregating them into like-operations so that like-work may be executed together in batches.

Monte Carlo transport functions lend themselves to asynchronous programming schemes, as it is intuitive to provide a function for each particle operation.
For example, Figure~\ref{fig:jitfunctions} shows a \texttt{move\_particle} function.
These functions can be ordered such that like operations get implemented in unison during runtime even if user defined control logic would dictate otherwise.
The end result of the computation is the same, but the order of execution on the processor has been optimized.
MC/DC calls Harmonize via Python bindings.
Harmonize has been shown to increase GPU performance by reducing thread divergence \cite{brax2023}.

\input{fig/cpuvgpu_func}

% description of GPU mode execution
Moving to compile and run Numba \texttt{jit}ed functions to the GPU requires making a few alterations to the kernels themselves.
An even-smaller subset of Python functions work in GPU-compiled code, with operations supported on Numba-CPU like \texttt{numpy.linalg.solve()} losing support.
Other operations may require API-specific calls, exposed by Numba commands.
For example, atomic operations are required to preserve the side-effects of individual threads acting on global memory (e.g., adding to a tally).
To allow for a mostly unified kernel base in MC/DC for both CPUs and GPUs, we track alternate function implementations registered through decorators.

Figure~\ref{fig:forcpuvgpu} shows how we implement alternate tally accumulation functions using \texttt{@for\_cpu} and \texttt{@for\_gpu} decorators.
Here \texttt{@for\_cpu} adds one to a value in an array, and since this is within a single MPI rank we can assume a thread-safe operation.
However, on the GPU this may result in a memory race condition requiring an \texttt{numba.cuda.atomic\_add} API call.
While this does increase complexity for a programmer implementing numerical methods, it is nowhere near the complexity that might be required to accomplish a similar implementation in a compiled language.

Most numerical methods development in MC/DC is done by editing pre-existing control flow (e.g., adding more operations or device functions to existing loops, adding more components to a data structure).
Once all alterations can compile and execute using Numba-CPU functionality and necessary API calls have been abstracted, MC/DC and Harmonize automatically compile and execute those extra commands on GPUs.
So, in most cases, methods developers do not need to interface with Harmonize commands or make any alterations to the GPU runtime, data management, or compilation techniques.

If more-significant alterations are required for a given numerical method, a developer may have to interface directly with Harmonize.
We have found that, for the majority of our work exploring new algorithms to date, Harmonize+Numba sufficiently abstracts the SIMT parallelism paradigm such that operations that work on the CPU side are generally supported on GPU with little effort from the methods developer.

% descirption of compilation + harmonize
To compile functions to GPU targets with Harmonize, Numba generates intermediate compiler representations (IRs, e.g., LLVM-IR or PTX) of Monte Carlo neutron transport kernels. 
Harmonize then ingests and links those IRs with the event-scheduling runtime.
MC/DC's documentation\footnote{\url{https://mcdc.readthedocs.io/en/dev/theory/gpu.html}} provides a more in-depth description how MC/DC and Harmonize are JIT compiled for given hardware.

When running MC/DC in GPU mode on an individual MPI thread, MC/DC+Harmonize is first JIT compiled, then  during initialization allocates device memory for the global array and moves this from the host (CPU) to the device (GPU).
Next, MC/DC's transport kernels are executed with Harmonize on the GPU until transport for a given collection of work is complete.
Communication between the GPU and CPU of the global variable may be required during transport for some simulation modes.
When transport is finished, the global variable moves back to the host for a final time, and the simulation completes.
% workflow description

Just as with CPU development, this abstraction strategy has some potential disadvantages.
%for most kernels between CPU and GPU runtime as compared to low-level language implementations
While MC/DC's software engineering structure allows for kernel portability between CPUs and GPUs significant time and effort can be lost in debugging, particularly for the data structures.
%Furthermore compiling device kernels from Numba Python with push compiler tool chains to their limits---often resulting in outright failure and undocumented behaviors.
%While the initial setting up of the workflows can be very difficult maintaining them is somewhat simpler
MC/DC only operates on GPUs using Harmonize.
Beyond its event scheduling and runtime capabilities, Harmonize allows us to ameliorate issues in Numba's GPU feature set.
For example, allocating and moving data from the CPU to GPU can only happen from Python code and cannot be done from Numba-compiled CPU kernels (requiring an \texttt{object\_mode} call).
In our initial implementations this required many copies of the global variable, which proved prohibitively costly for larger problems.
Using API calls elevated through Harmonize instead of Numba fixes this issue, requiring only two copies of the data, and the data can be accessed from both Numba-compiled CPU and GPU kernels.
In addition, when extending GPU operability to other vendors (namely AMD GPU support), Harmonize allows us to elevate non-implemented Numba API calls to the MC/DC Python interface.
For example, the Numba-HIP package\footnote{\url{https://github.com/ROCm/numba-hip}} does not currently support atomic operations on vectors.
Harmonize provides a clear path to elevate HIP-C\texttt{++} functions into Python for use in MC/DC.

%While MC/DC's software engineering structure allows for kernel portability between CPUs and GPUs significant time and effort can be lost in debugging, particularly for the data structures.

For GPU development, the portability and performance enabled by MC/DC's software engineering structure increases the difficultly of implementation for the workflow developer who actually interfaces with Numba and Harmonize.
Our hope is that the investment made by the workflow developers is compounded with rapid development of more numerical methods.

\section{Performance}

\begin{figure}[h]
    \centerline{
    \includegraphics[width=.4 \textwidth]{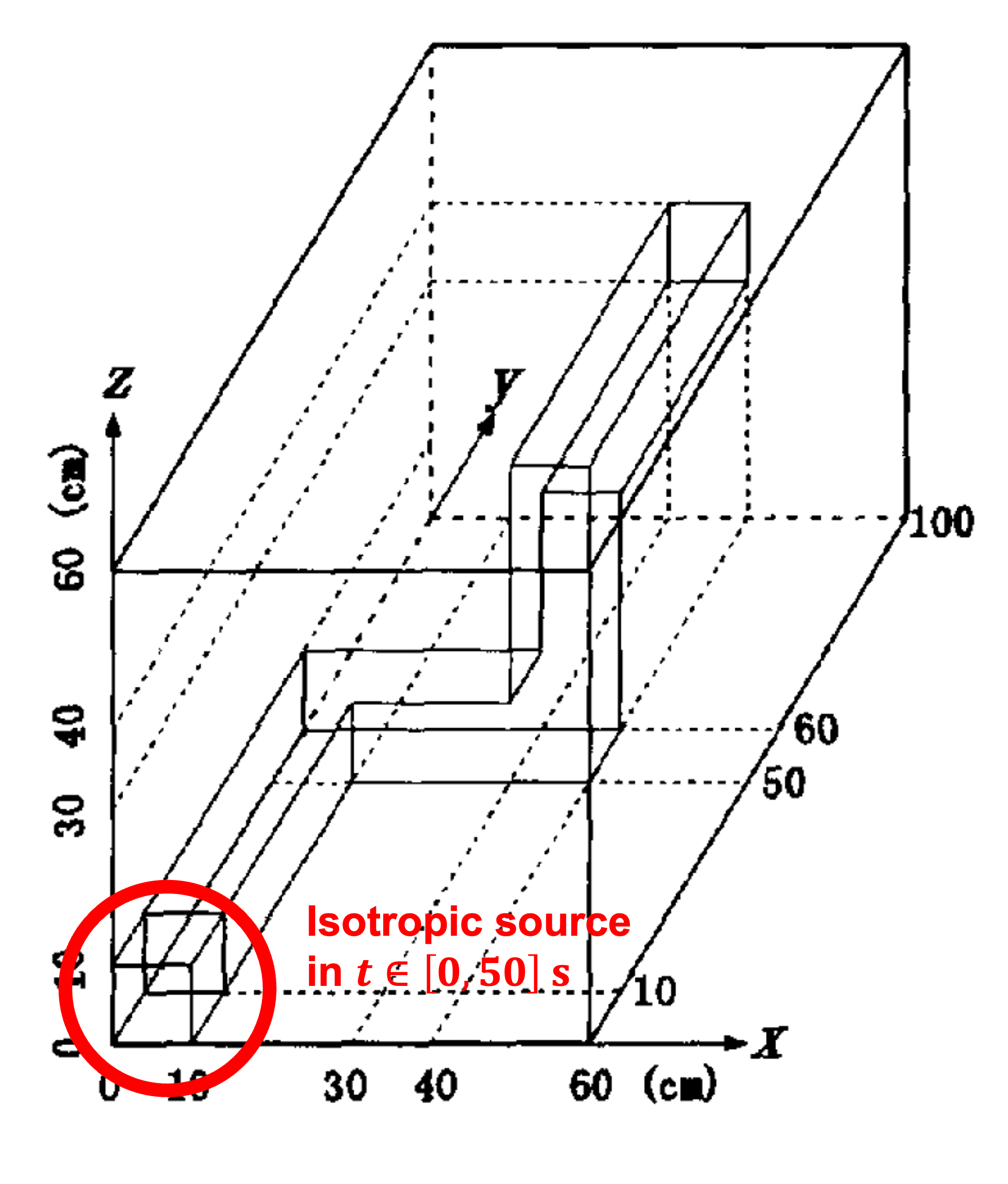}
    } 
    \caption{Kobayashi problem schematic.}
    \label{koby-problem-def}
\end{figure}

\begin{figure*}[h]
    \centerline{
    \includegraphics[width=\textwidth]{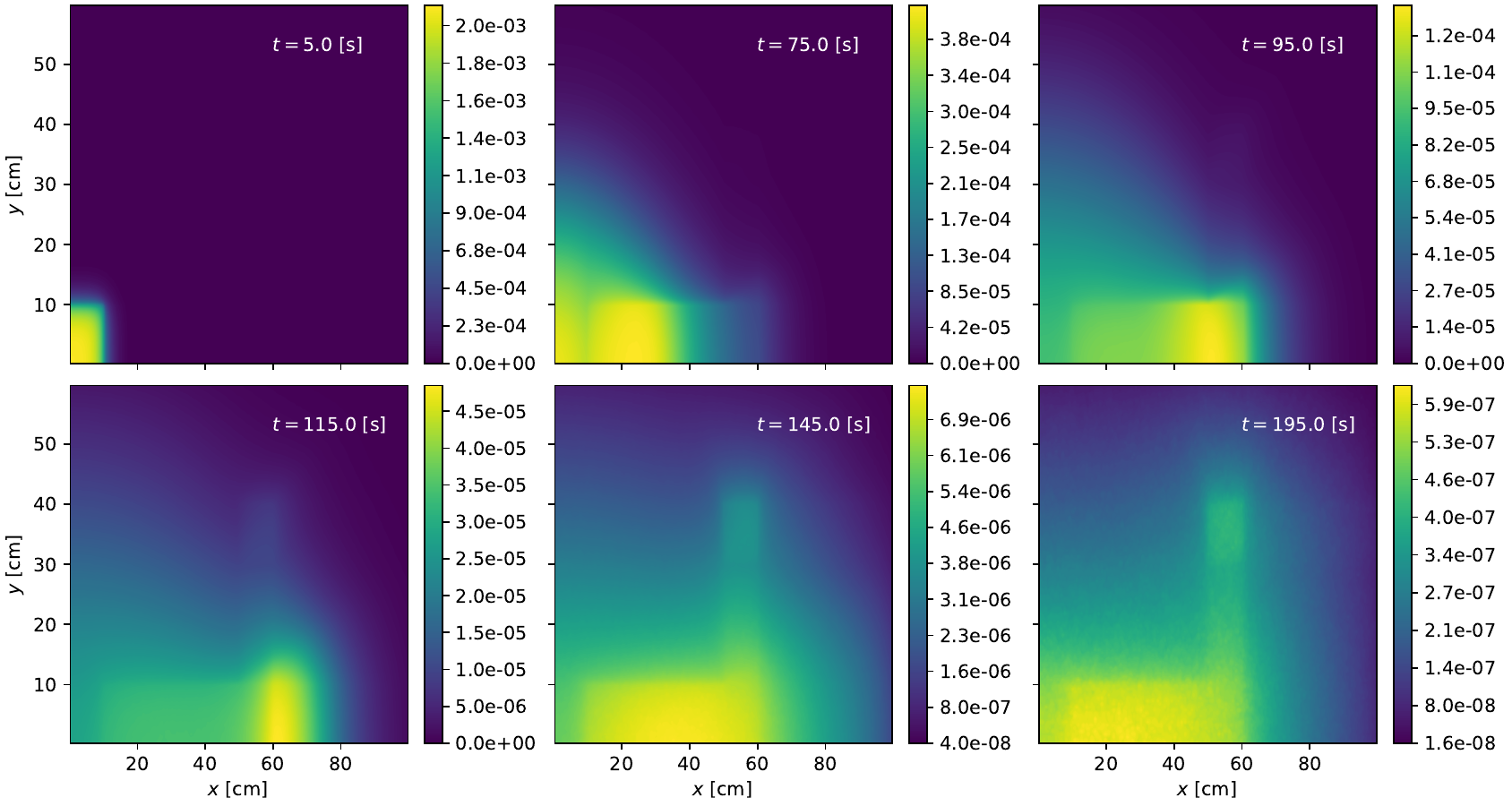}
    } 
    \caption{Time and space averaged scalar flux solution to the Kobayashi problem run with $1\times 10^{9}$ particle histories at various points in time.}
    \label{koby-results}
\end{figure*}

% transient runtime strong scaling mcdc v openmc maybe v shift?
To examine the performance of MC/DC we use a time-dependent version of the one-group Kobayashi dog-leg void-duct problem \cite{Kobayashi2001, variansyah_mc23_mcdc}.
Figure~\ref{koby-problem-def} shows the void duct and the location of the neutron source at the opening of the duct.
The initial condition is zero flux everywhere.
Radiation quickly moves through the void and then penetrates the walls of the problem, slowly dissipating through time.
Figure~\ref{koby-results} shows the duct clearly with the scalar flux solution at various points in time.

We solved the Kobyashi problem on HPC systems available at Lawrence Livermore National Laboratory (LLNL): the Dane and Lassen machines.
Dane is a CPU-only system with dual-socket Intel Xeon Sapphire Rapids CPUs, each with 56 cores for a total of 112 per node.
Lassen has four Nvidia Tesla V100s and two IBM Power 9 CPUs per node.
To contrast MC/DC on the CPU against a traditionally developed and compiled code, we will compare performance to another Monte Carlo neutron transport code, OpenMC\footnote{\url{https://github.com/openmc-dev/openmc}} \cite{romano_openmc_nodate} (an open-source code written in C\texttt{++}).
We added time-dependent functionality to OpenMC\footnote{\url{https://github.com/CEMeNT-PSAAP/openmc/tree/transient}} so that the same algorithm is implemented in both codes for the Kobyashi problem.

\begin{figure*}[h]
    \centerline{
    \includegraphics[width=1.1\textwidth]{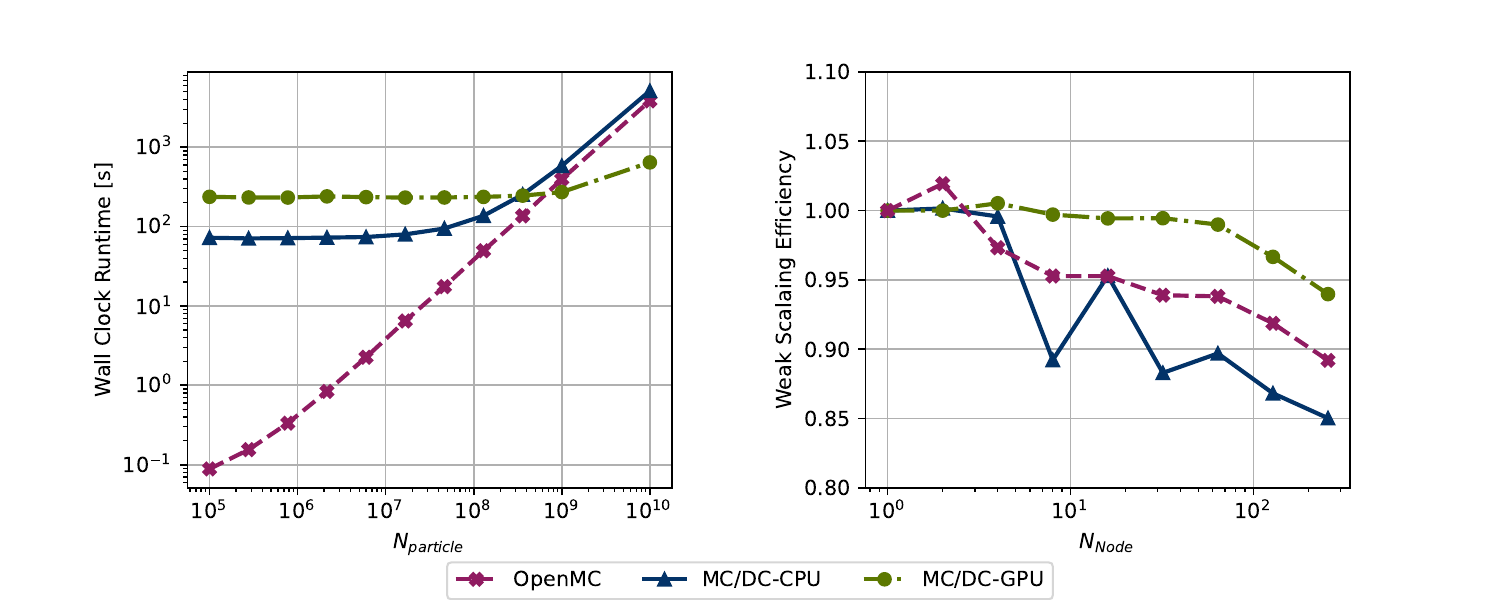}
    }
    \caption{Left: Wall-clock runtime of the Kobyashi problem over particle counts. 
    Right: Weak scaling efficiency as a function of node count for the Kobyashi problem on Dane (CPU) and Lassen (GPU).}
    \label{performance_results}
\end{figure*}

Figure~\ref{performance_results} at left shows the wall-clock runtime of OpenMC (112 MPI threads), MC/DC-CPU (112 MPI threads), and MC/DC-GPU (four MPI threads) using all available resources of a given node type.
Both MC/DC runs are JIT compiled, which means compiling consumes a considerable amount of wall-clock runtime for even small problems (about \SI{70}{\s} and \SI{140}{\s} for CPU and GPU targets, respectively).
For small particle counts, actual compute time is small relative to compile time, so both MC/DC lines are flat until enough work saturates the computational power of a given resource---around \num{e8} particles for MC/DC-CPU and \num{e9} for MC/DC-GPU.
At full saturation (\num{e10} particles) MC/DC-CPU runs about 22\% slower than OpenMC, while MC/DC-GPU is 8$\times$ faster than MC/DC-CPU and 6$\times$ faster than OpenMC.

OpenMC displays superior performance at smaller particle counts due to it being a fully compiled code.
GPU profiling for MC/DC shows that
\texttt{memalloc} and \texttt{memcopy} CUDA API calls 
occupies 2.2\% of runtime (\SI{32.8}{\s} out of \SI{1500}{\s})
when running \num{e10} particles on one Lassen GPU for the Kobyashi problem.
At \num{e9} particles, GPU memory commands account for 11.8\% of runtime (\SI{18.0}{\s} out of \SI{147.0}{\s}).

Figure~\ref{performance_results} at right shows weak-scaling performance (\num{e10} particles per node) with the Kobyashi problem for MC/DC-CPU (Dane), OpenMC (Dane), and MC/DC-GPU (Lassen).
Each node is using all available compute resources for a given calculation (e.g., four nodes is 480 CPU cores and 16 GPUs on Dane and Lassen, respectively).
MC/DC-CPU shows a minimum efficiency of \num{0.85} at 256 nodes while OpenMC only falls to \num{0.89}.
OpenMC supports shared-memory parallelism (using OpenMP) but these calculations only use domain-replicated MPI.
MC/DC-GPU shows the best weak-scaling efficiency for this problem, decreasing only to \num{0.95} at 256 nodes.

\section{Discussion, Conclusions, and Future Work}

% perfomrance conclusions
%\former{Adding three words.}{editor}
Monte Carlo/Dynamic Code (MC/DC) is a Monte Carlo neutron transport code that targets modern HPC architectures with CPUs and GPUs.
Our performance results demonstrate that MC/DC's structure using a Python + Numba + MPI + Harmonize scheme can produce similar performance to other Monte Carlo neutron transport solvers.
After JIT compilation overhead, MC/DC performs similarly to traditionally compiled production code on a single node for a transient problem of interest.
MC/DC exhibits similar weak scaling on CPUs and superior weak scaling on GPUs up to 256 nodes of a given HPC, compared with a CPU-only production code.

Developing using Numba for CPU targets can be as difficult as developing in low-level languages for the complicated algorithms we implement. 
This agrees with previously published analysis \cite{KailasaSrinath2022PAEi}.
We found developing the necessary time-dependent features to model the Kobyashi problem in OpenMC to be about as difficult as making changes within MC/DC.
For our application, anything gained when using a high-level-language is lost in time and effort spent circumventing unsupported operations and debugging. 
However, the implementation in OpenMC remains CPU-only, while for MC/DC it took little effort to go from a working CPU implementation to something operating and highly-performing on GPUs.
Of course, we use our own specialized event-scheduling library to do this---but Numba allows us to construct a Python-based portability framework fit to our numerical method with the added benefit of unifying our high-level glue language and kernel-production language.

Over the duration of developing MC/DC (starting in 2021) we have seen many improvements to Numba.
Compiler error reporting continues to improve (especially in Numba versions 0.59.0$+$), the number of supported operations have grown, and Numba has been extended to additional accelerators like AMD and Intel\footnote{\url{https://github.com/IntelPython/numba-dpex}} GPUs.
We have found that the Numba development team fosters a supportive community that is approachable and responsive to questions, comments, and concerns.
We believe that as Numba matures we will continue to see performance and development improvements.

%When we started this project we had to manually patch Numba in many locations, as of Numba v0.60.0 patching is no-longer required.
%Only time will tell if Numba sees continued support for new features and packages but we are hopeful that future development will continue.}{editor}
%When looking at more general conclusions we can draw from our experience is that Numba+Python is a compelling development scheme to enable portability between CPUs and GPUs.
%However (very much depending on a case-by-case basis) much more may be required to abstract the differences GPUs to computing GPUs require.

% future work CPUs
Work in MC/DC is ongoing.
We are continually exploring novel variance reduction and hybrid Monte Carlo techniques, and adding new functionality.
For GPU development specifically we are currently investigating use of unified memory between the CPU and GPU as well as extending support to Intel GPUs.
We will continue to improve MC/DC, making it a portable application for rapid methods development enabled by Python and Numba.

\section{Acknowledgments}
The authors thank the Numba development team for support using the Numba compiler as well as Damon McDougall and Dominic Etienne Charrier from Advanced Micro Devices for support using Numba-HIP and ROCm compilers.
The authors thank the high performance computing staff at Lawrence Livermore National Laboratory for continued support using the Dane and Lassen machines.

%The authors thank William Godoy from Oak Ridge National Laboratory, blah from Lawrence Berkeley National Laboratory and blah from blah for useful conversations about domain specific languages and rapid methods development for HPCs.

This work was supported by the Center for Exascale Monte-Carlo Neutron Transport (CEMeNT) a PSAAP-III project funded by the Department of Energy, grant number: DE-NA003967.
% the actual text of the article is here for ease of porting from format to format

%% file: fig/cpuvgpu_func.tex
\begin{figure}
\begin{lstlisting}[language=Python]
from numba import cuda

@for_cpu
def add(array, value, idx):
    array[idx] += value

@for_gpu
def add(array, value, idx):
    cuda.atomic_add(array, value, idx)

def tally_collision_event(mcdc, part):
    id = loc2index(part)
    add(mcdc.col_tally, part.v, id)
\end{lstlisting}
\caption{Example of GPU and CPU specific API calls as defined in MC/DC and their use in a collision tally function.}
\label{fig:forcpuvgpu}

\end{figure}